\documentclass{article}
\usepackage{arxiv}

\usepackage{graphicx}
\usepackage{cite}
\usepackage{amsmath}
\usepackage{url}
\usepackage{color}
\usepackage{multirow}
\usepackage{array}

\begin{document}
%
\title{Security of HyperLogLog (HLL) Cardinality Estimation: Vulnerabilities and Protection}
%
%
%


\author{
  Pedro Reviriego \\
  Universidad Carlos III de Madrid \\ 
  Legan\'es 28911, Madrid, Spain \\ 
  email: {\tt\small revirieg@it.uc3m.es}  \\
   \And
  Daniel Ting \\ 
  Tableau Software \\ 
  Seattle, Washington, USA \\
  email: {\tt\small dting@tableau.com} \\
}

\maketitle


\begin{abstract}

Count distinct or cardinality estimates are widely used in network monitoring for security. They can be used, for example, to detect the malware spread, network scans, or a denial of service attack. There are many algorithms to estimate cardinality. Among those, HyperLogLog (HLL) has been one of the most widely adopted. HLL is simple, provides good cardinality estimates over a wide range of values, requires a small amount of memory, and allows merging of estimates from different sources. However, as HLL is increasingly used to detect attacks, it can itself become the target of attackers that want to avoid being detected. To the best of our knowledge, the security of HLL has not been studied before. In this letter, we take an initial step in its study by first exposing a vulnerability of HLL that allows an attacker to manipulate its estimate. This shows the importance of designing secure HLL implementations. In the second part of the letter, we propose an efficient protection technique to detect and avoid the HLL manipulation. The results presented strongly suggest that the security of HLL should be further studied given that it is widely adopted in many networking and computing applications.

\end{abstract}


%

\section{Introduction}

Network monitoring is a key element for security \cite{NSM}. Monitoring can be done at many levels. For example, each packet's payload can be inspected to detect malicious contents or the header can be analyzed to identify packets coming from suspicious sources. Monitoring can also be done at a coarser granularity. For example, by looking only at the link loads. The number of connections for each host or network can also be used to detect attacks or abnormal activity \cite{Cardinality_Detection_Infocom}. This can be done by storing the connections in a table and, for each new packet, checking if the connection is already in the table. This, however, requires a significant amount of memory if the number of connections or networks to monitor is large \cite{Snare}.  Counting the number of different connections is basically identifying the number of distinct elements (connections) in a set (packets) which has been widely studied \cite{Cardinality}. This problem is also known as cardinality estimation and many algorithms have been proposed over the years. For example, Linear Probabilistic Counting Arrays (LPCAs) are efficient when the expected range of values is small and known in advance \cite{Bitmaps}. Algorithms that can cover a wide range of cardinalities have also been developed, like for example the HyperLogLog (HLL) \cite{HLL} which has been widely adopted \cite{HLL_Google}.

HyperLogLog has a number of advantages that make it attractive. Compared to a bitmap based sketch like LPCA that can become saturated, HLL provides good estimates over a large range of cardinality values. 
It does so with a small memory footprint, requiring only $\approx 5 k$ bits to achieve errors of $100/\sqrt{k}$ percent.
Furthermore, HLL sketches can be merged to compute the cardinality of unions. This is useful in network applications as measurements may be taken on different nodes and combined to obtain better and wider estimates \cite{Worm_cardinality}. In fact, HLL is widely used in networking for example to detect malicious worm activity \cite{Worm_cardinality}, to detect nodes with a large number of connections \cite{HLL_Networks} or networks scans made by attackers \cite{HLL_Switches}. 

A potential issue when using any detection mechanism is that an attacker can try to evade detection. Therefore, since HLL is widely used for network monitoring, it is interesting to consider what an attacker can do to evade detection based on HLL cardinality estimation. To the  authors' best knowledge, there is no previous work on the security of HLL. The closest works focus on the privacy implications of HLL \cite{Cardinality_Privacy} or on attacks on data structures like Bloom filters or the Count-Min sketch \cite{BF_attacks}. 

In this paper, for the first time, the security of HLL is considered and two interesting contributions are made. The first one is to show that an attacker can easily manipulate the HLL estimate with no knowledge of its implementation if he can add elements and access the HLL estimates. Uncovering this vulnerability would help designers to protect their HLL implementations and would also raise awareness on the importance of security in HLL. The second contribution is an efficient protection scheme that detects and avoids HLL manipulation and preserves the crucial mergeability property of HLL sketches. The results presented in this paper are also intended to foster further research to study the security of HLL.

The rest of the paper is organized as follows. Section 2 gives a brief overview of HLL and introduces the notation used in the paper. Section 3 discusses the security of HLL, first describing the adversarial models considered and then presenting the proposed attack to manipulate the HLL estimate. Then in section 4, the techniques to detect and avoid the attack on HLL are presented and their practical application is discussed in section 5. The paper ends in section 6 with the conclusion and some ideas for future work.

\section{HyperLogLog}

\begin{figure}[h]
  \centering
  \includegraphics[scale=0.4]{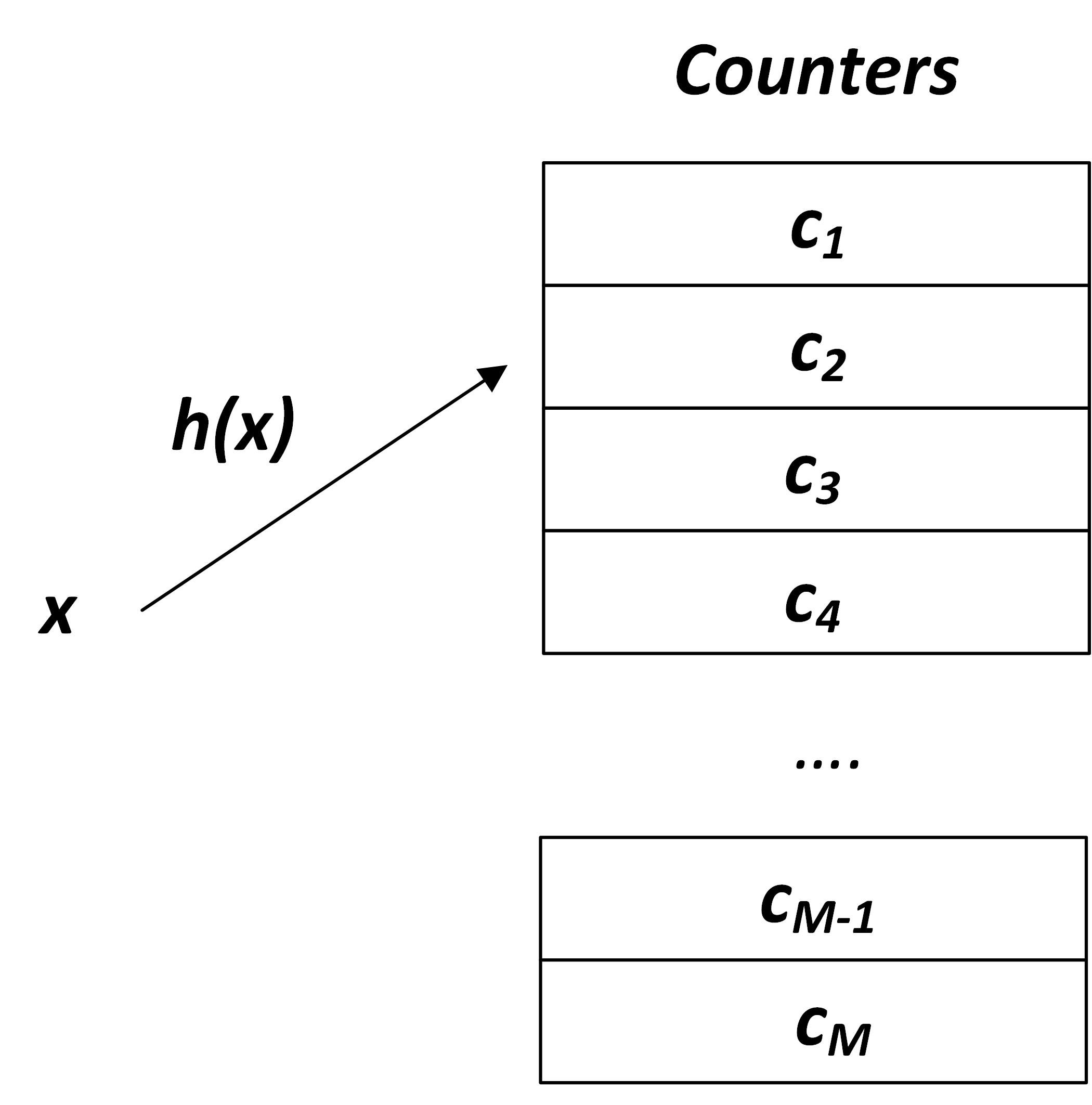}\\
  \caption{Diagram of the structure used in HLL 
  }\label{FigHLL}
\end{figure}

The HyperLogLog (HLL) algorithm \cite{HLL} is a popular and highly efficient data sketch for estimating a data stream's cardinality, in other words, the number of unique elements in the data stream. It is composed of an array of $M$ counters $c_1, \ldots, c_m$ that each contain a very small integer value that can typically be stored in 5 bits. Elements $x$ are first mapped to one counter using a hash function $h(x)$. By using a hash function, a duplicate element is always mapped to the same counter. Figure \ref{FigHLL} illustrates the structure used in HLL. 
 
\begin{figure}[h]
  \centering
  \includegraphics[scale=0.4]{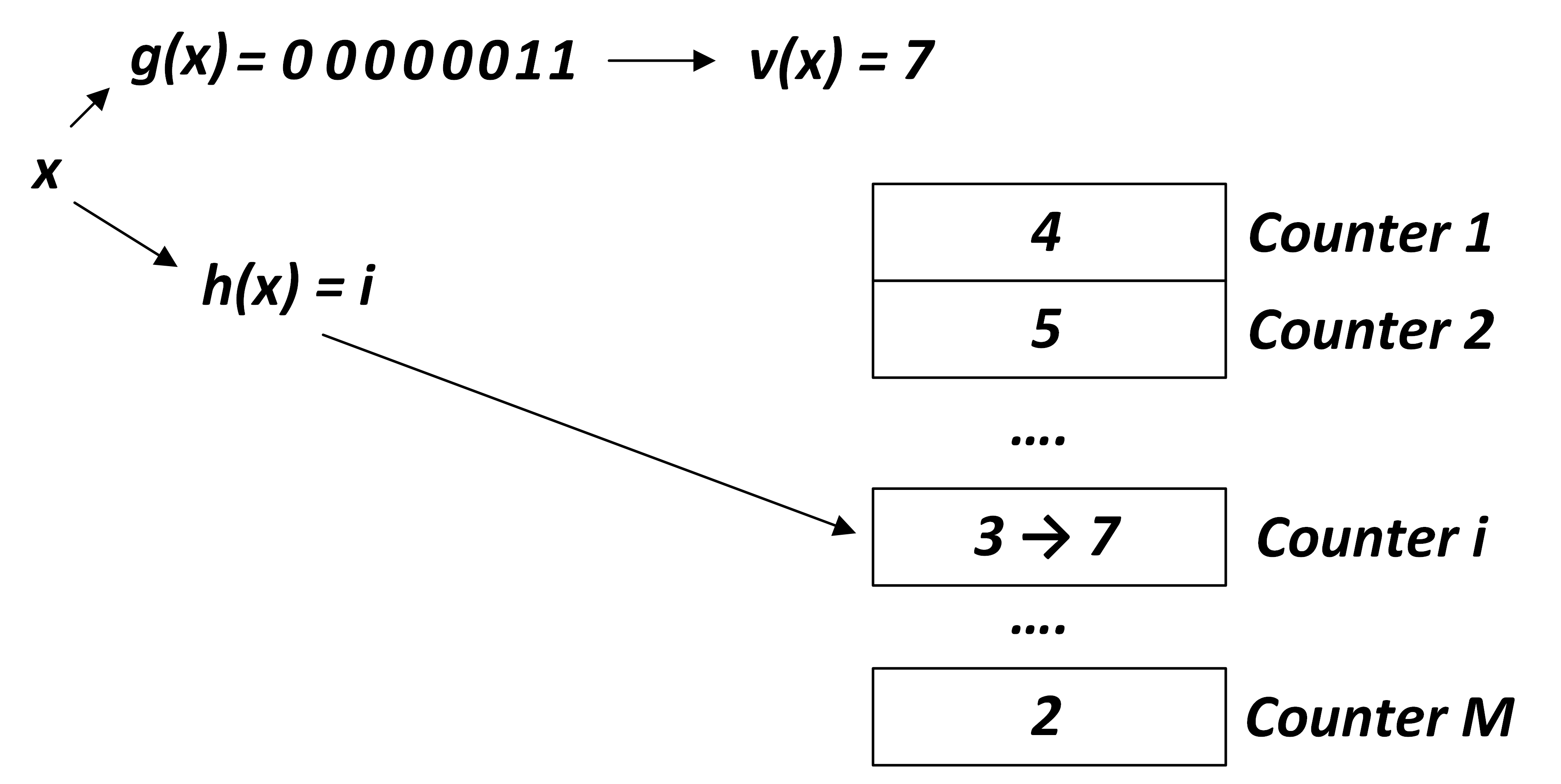}\\
  \caption{Example of a counter update in HLL}\label{FigHLL2}
\end{figure}

To update a counter, another hash function $g(x)$ is used to generate a value.
This value $v(x)$ is one plus the number of leading zeros in the bit representation of $g(x)$.
Each counter then keeps the maximum value for all elements mapped to that counter.
For example, consider a counter $c_i$ that two elements $y$ and $z$ are mapped to and suppose $g(y)= 01011001$ and $g(z)= 00010101$. This gives $v(y) = 2$ and $v(z) = 4$, and this counter stores $c_i = v(z) = 4$. 
The update of a counter in HLL is illustrated in Figure \ref{FigHLL2}. In this case, element $x$ is inserted by first selecting the counter with $h(x)$. Then $v(x)$ is computed based on $g(x)$ and a value of 7 is obtained. Since this is larger than the value of 3 stored in the counter, the counter value is updated to 7. If $x$ is inserted again later, it will not increment the counter value. The idea of these counters is that the probability of having a value with $t_z$ leading zeros is related to the number of elements mapped to the counter in an exponential way. This enables the counters to cover a wide range of cardinalities using only a few bits.     

The cardinality in HLL is estimated by first computing the harmonic mean of the counter contents as follows:

\begin{equation}
  Z = \frac{1}{\sum_{i=1}^{M} 2^{-c_i}}, 
\end{equation}

\noindent where $c_i$ is the value of the $i^{th}$ counter.   

The cardinality estimate is obtained by multiplying by a constant factor given in \cite{HLL} that can depend on the number of counters $M$: 

\begin{equation}
\label{eqn:harmonic mean}
  C = {\alpha}_M \cdot M^2 \cdot Z
\end{equation}

For large cardinalities, the error of the HLL estimate is approximately $\frac{1.04 n}{\sqrt{M}}$
where $n$ is the true cardinality. Therefore, a desired accuracy can be obtained by adequately selecting the number of counters $M$. An interesting feature of HLLs is that they can be merged in order to compute the cardinality of the union of data streams. To merge a collection of HLL sketches constructed using the hash functions $g$ and $h$, for each counter position, select the maximum counter value across all the sketches. The resulting HLL sketch is identical to one computed on all data streams concatenated together. In network monitoring, merging is useful to combine measurements from different nodes.    

\section{Vulnerabilities of HyperLogLog}

This section first presents two models for the attacker. These depend on the information the attacker has about the HLL implementation. We then present schemes to manipulate the HLL estimates for both attacker models.

\subsection{Adversarial Models}

The best case for an attacker would be to know the hash functions used in the HLL implementation: $h(x)$ and $g(x)$. This can be a realistic assumption if HLL is implemented using open source software libraries or commonly used hash functions and default seeds. This first model will be denoted as $M1$ in the rest of the paper.

A much weaker assumption is that the attacker can insert elements into the sketch and check the cardinality estimate. This can easily be done if the attacker knows which device is used to do the monitoring\footnote{He just needs to get the same device and test it.}. This second model will be denoted as $M2$ in the rest of the paper.

\subsection{Manipulating HLL Estimates}

To evade HLL detection, an attacker needs to manipulate the HLL estimate so that it is much lower than the real number of distinct elements in the set. That would allow, for example, a large number of connections to be established without triggering an alert. Consider the case where HLL monitors the number of flows in a network and each flow is identified by the 5-tuple of source and destination IP addresses and transport ports plus the protocol. The attacker has a set of flows $A$ with cardinality $C_a$ and wants that they do not increment the cardinality estimate of HLL $C_{HLL}$. This would be achieved if for all the flows $a$ in $A$, the hash function $g(a)$ starts with a '1'. Equivalently, the counter value for the flow is $v(a) = 1$. 
If that is the case, the flows in $A$ would not increment any non-zero HLL counters, and it is unlikely to affect the HLL estimates at all for even moderate cardinalities. This is since an HLL sketch with $M$ counters is expected to have no non-zero counters after encountering $M \log M$ distinct items.

Let us start by assuming an $M1$ attacker that knows the function $g(x)$. Fixing all other values, the attacker can modify the source port value to find an element $a$ such that $g(a)$ starts with a '1'. Generating many of such elements is easy as half of the values start with '1' in expectation. Therefore, a set of flows $A$ that would go undetected can easily be created by the attacker with arbitrary source and destination IP addresses, protocol and destination port values. The same reasoning applies even if detection is based only on one field, for example the source IP address. In that case, if an attacker controls a number of nodes, roughly half or more of these nodes will not result in any change in the HLL sketch. This is worrying as it shows that the attacker can easily generate a large set $A$ that will not be detected. 

Even when considering an $M2$ attacker that does not have access to $g(x)$, he can still build a set $A$ that has estimated cardinality much smaller than the true cardinality $|A|$. Assume that the attacker can insert into the HLL sketch and check the estimate\footnote{This can be done by using a device with the same HLL implementation.}. Then, he can insert elements on a HLL and check if they increase the cardinality value or not. Elements that do not increase the estimate can be added to $A$ as they are less likely to increase the HLL estimate. In \cite{Cardinality_Privacy}, it was shown that a similar procedure could be used to extract information about whether an element was part of an HLL. In section 5, the construction of a set $A$ for an existing HLL implementation using those techniques is presented to demonstrate the feasibility of the attack. 

In the previous discussion, the attacker tries to reduce the HLL estimate to avoid detection. Another attack artificially increases the HLL estimate, for example, to create a flood of false positives that overwhelm the defender's ability to assess if warnings are genuine threats. In this case, creating the set $A$ is more challenging as we need to find elements whose $g(x)$ starts with a large number of zeros but it still can be done. In the rest of the paper we focus only on an attacker that is trying to evade detection by reducing the HLL estimate.    

\section{Protecting HyperLogLog}

From the previous section, it is apparent that even an attacker with no access to the implementation details of HLL ($M2$ attacker model) can evade detection if he can test insertion and estimation. This is worrying as HLL is widely used in network monitoring. Therefore, it is of interest to consider techniques to protect against attackers evading HLL. In this section, existing protection techniques are discussed, and a new protection technique is presented.  

\subsection{Existing techniques}

To protect HLL against attacks, 
random, one-time use salts can be used in the hash functions for every instance of an HLL sketch \cite{Cardinality_Privacy}. That is, instead of using hashes $h(x), g(x)$, the hashes can be written as $h(x,s),g(x,s)$ with $s$ being a random number, alternatively $s$ can be used as a parameter to configure the hash. An attacker would not be able to infer if an element $x$ would modify the estimate of an HLL unless it has access to that particular HLL instance's salt.
Therefore, the use of a salt would effectively protect the HLL unless the attacker has gained access to the monitoring device. But when that is the case, he can probably evade detection in other ways and evading the HLL would not be an issue.  
The use of a salt has, however, a major drawback, HLLs can no longer be merged unless they use the same value of $s$ which requires sharing $s$ as a secret among all nodes \cite{Cardinality_Privacy}. This can be an important limitation as in network monitoring, merging is useful for aggregating results from different nodes.

To preserve the ability of merging HLLs, another potential approach is to add "noise" to the HLL estimate, for example by rounding it \cite{Cardinality_Privacy}. This would make harder for the attacker to identify which elements modify the cardinality estimation. 
However, it seems that the attacker could still get that information \cite{Cardinality_Privacy}. This can be done by testing different combinations of elements repeatedly.  Thus, although rounding or adding noise to the estimate makes evasion harder, it does not prevent it.

\subsection{Salted and not Salted Protection (SNS)}

\begin{figure}[h]
  \centering
  \includegraphics[scale=0.6]{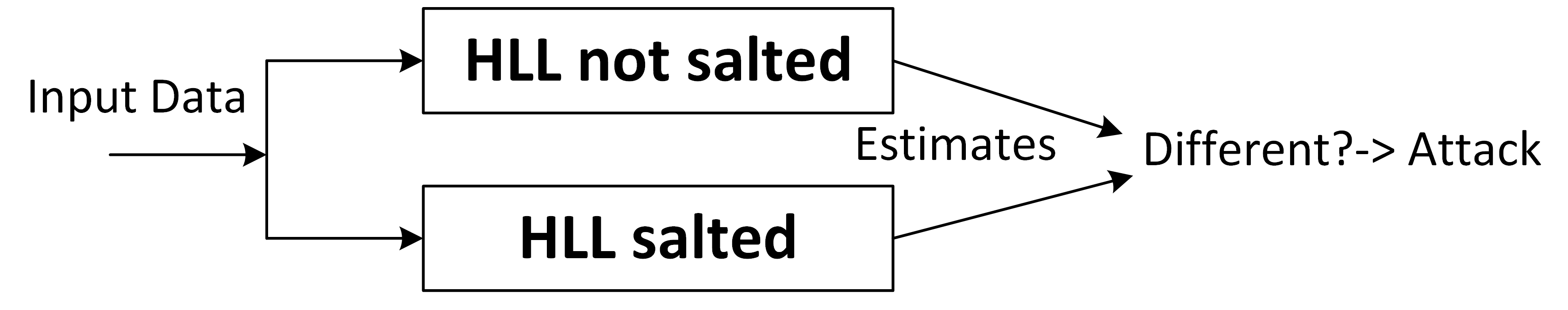}\\
  \caption{Diagram of the structure used in the proposed Salted not Salted scheme 
  }\label{FigSNS}
\end{figure}

To avoid evasion while preserving the ability to merge HLLs, in the following we propose the SNS protection technique based on the combination of two HLLs, one salted and one not salted. The idea is to use the redundancy to detect evasion attacks while preserving mergeability when there is no attack. The overall approach is illustrated in Figure \ref{FigSNS}. Both HLLs are updated simultaneously and their estimates are compared. If their difference is larger than expected, then we infer that an attempt has been made to attack the HLL. If the estimates are similar, then the not salted HLL can be used for merging.  This scheme in addition to avoiding evasion, detects when an attacker is trying to evade the HLL.  
 
In detail, the structure used by SNS is composed of two HLLs, the first one is randomly salted and the second one not. Let us consider that they have $M_s$ and $M_{ns}$ counters respectively and that their estimates are $C_s$ and $C_{ns}$. Then, each time we check the cardinality we also check if the difference between ${C_s}$ and $C_{ns}$ is larger than a given fraction $d_t$ of $C_{s}$. If so, we consider that we are under attack and we can only use $C_{s}$ as a valid estimate.  Therefore, the selection of $d_t$ is critical to ensure reliable detection of attacks while minimizing false positives. 

The cardinality estimate of HLL is approximately normally distributed with mean equal to the true cardinality $C$ and standard deviation $1.04 \cdot C \cdot \sqrt{\frac{1}{M}}$. This is since the denominator in the harmonic mean in equation \ref{eqn:harmonic mean} is asymptotically normal by the central limit theorem. The cardinality estimate also asymptotically normally distributed by the delta method \cite{casellaberger} since it is a continuously differentiable function of this harmonic mean in a neighborhood away from $0$. The standard deviation for large cardinalities is given by \cite{HLL}. For small to moderate cardinalities \cite{tingManyHLL} provides accurate estimators for the standard deviation of an improved cardinality estimator. 

Therefore, a one sided difference in means test can be used to test if there is no attacker. Two HLL sketches fed the same input would return estimates whose difference $C_{ns} - C_{s}$ is normally distributed with zero mean and standard deviation $\sigma  = 1.04\; C \sqrt{\frac{1}{M_{ns}}+\frac{1}{M_{s}}}$. The ratio $(C_{ns} - C_{s}) / C_s$ will also be approximately normally distributed with mean $0$ and standard deviation $\sigma$.
This allows us to set the detection threshold $d_t$ to achieve a desired false positive probability
for the test $(C_{ns} - C_{s}) / C_s < -d_t$.
The false positive probability is given by $\Phi(-d_t / \sigma)$
where $\Phi(x) = P(Z \leq x)$ is the cumulative distribution function for the $Normal(0,1)$ random variable $Z$. For example, setting $d_t = 3 \cdot \sigma$ would yield an approximately $0.1\%$ false positive probability while $d_t = 6 \cdot \sigma$ yields a minuscule false positive probability of approximately $10^{-9}$.

The overhead introduced by the proposed SNS scheme is basically the need to keep two HLLs which implies doubling the number of operations in terms of hash computations and memory accesses. The memory for the counters would also be doubled when the sketches are equally sized so $M_{ns} = M_{s}$. It is also possible to use a smaller $M_{s}$ at the cost of a less accurate verification procedure and larger threshold $d_t$.    

When there is no attack and no sketches are merged, the estimates $C_s$ and $C_{ns}$ are independent and can be averaged to obtain a more accurate estimate.
In this case, there is effectively no space overhead since the standard deviation of the averaged estimate, weighted by sketch size, is the same as that of a sketch using $M_s + M_{ns}$ counters.

\section{Practical Application}

In this section, the proposed attack on HLL and the SNS protection are illustrated from a practical perspective. First, the construction of a large set of elements that gives a low cardinality estimation when using a publicly available tool is presented and then the practical configuration of SNS is described.

\subsection{Manipulating HLL estimates}

To show the feasibility of the proposed attack, we exploit the cardinality estimation in Redis \cite{Redis}. Redis is a network based publicly available key-value store that includes cardinality estimation based on HLL. To generate an attack, start with 250,000 distinct elements. These are sequentially added to a new HLL sketch. After each insertion the cardinality is checked and the elements which do not increment the HLL estimate are retained. This procedure is repeated three times starting with the last set of retained elements. This filters out elements that increment the HLL estimate. The end result is a set with 74,390 distinct elements that gives an HLL estimate of only 15,780 or roughly a five-fold reduction from the true cardinality. This clearly shows that an $M2$ attacker can easily craft a set $A$ that will bypass HLL detection. The procedure can be refined to further reduce the HLL estimate.

\subsection{SNS protection}

\begin{figure}[h]
  \centering
  \includegraphics[scale=0.46]{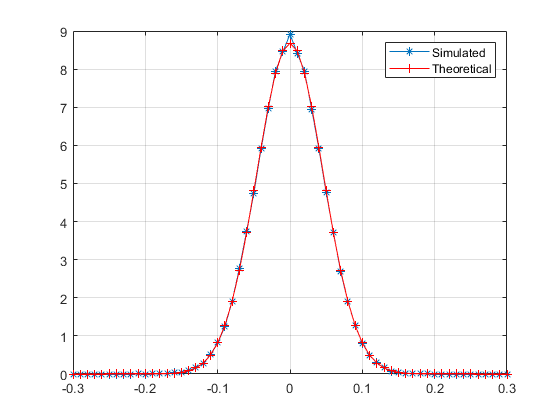}\\
  \caption{Probability distribution function of the difference of ${C_s}$ and $C_{ns}$ normalized by $C$ with $M_{ns} = M_{s} = 1024$.
  }\label{FigPDF}
\end{figure}

To describe the use of SNS, consider an implementation with $M_{ns} = M_{s} = 1024$. The value of $\sigma$ for this configuration would be approximately 0.046 and thus setting $d_t = 5 \cdot \sigma$ gives a value of 0.23. Therefore, any deviation between ${C_s}$ and $C_{ns}$ of more than 23\% would be a sign that there is an evasion attempt.  To validate the analysis, the two HLL have been simulated and the difference  ${C_s}$ and $C_{ns}$ has been measured when $C = 100000$. The results are shown in Figure \ref{FigPDF} and compared with the theoretical distribution. It can be observed that they match very well. This example shows how  the proposed method can be used to detect attacks with a very low number of false positives in a practical configuration.

\section{Conclusion}

This paper has considered the security of HyperLogLog (HLL) cardinality estimation. The analysis shows that an attacker can easily manipulate the HLL estimates even if he has no knowledge of the HLL implementation details. Traditional protection schemes like adding a salt to the hash computations of the HLL have a major disadvantage as the HLLs can no longer be merged. This is an issue for network monitoring as merging HLLs from different nodes provides valuable information. To avoid that issue, a new protection scheme has been proposed. The idea is to use both a salted and a not salted HLL on each node. When their estimates are significantly different, then an attacker is trying to manipulate the HLL estimates. The proposed salted and not salted (SNS) scheme can not only avoid manipulation, it also detects manipulation attempts. The applicability of the SNS scheme has been illustrated for a practical configuration showing how it can detect manipulation attempts with very low false positive probability. Therefore, it can be an interesting approach to protect HLLs from evasion when mergeability needs to be preserved.

\end{document}